\documentclass[sigconf,nonacm]{acmart}
\usepackage{fancyhdr}
\fancypagestyle{preprint}{%
  \fancyhf{}%
  \fancyhead[C]{\small
    Pre-print -- Extended version of the poster paper accepted at the 41st ACM/SIGAPP Symposium on Applied Computing (SAC) Smarter Engineering - Building AI and Building with AI (SEAI)~2026%
  }%
  \fancyfoot[C]{\thepage}%
}
\usepackage[T1]{fontenc}
\usepackage{graphicx}
\usepackage{makecell}
\usepackage{changepage}
\usepackage{booktabs}
\usepackage{amsmath}
\usepackage{listings}
\usepackage{subcaption}
\usepackage{graphicx}
\usepackage{xcolor}

\definecolor{pblue}{rgb}{0.13,0.13,1}
\definecolor{pgreen}{rgb}{0,0.5,0}
\definecolor{pred}{rgb}{0.9,0,0}
\definecolor{pgrey}{rgb}{0.46,0.45,0.48}
 \lstdefinelanguage{PrettyJava}
{ 
  language=Java,
  showspaces=false,
  showtabs=false,
  breaklines=true,
  showstringspaces=false,
  breakatwhitespace=true,
  commentstyle=\color{pgreen},
  keywordstyle=\color{pblue},
  basicstyle=\fontsize{5}{5} \ttfamily,
  moredelim=[is][\bfseries\textcolor{pgreen}]{/+}{+/},
  moredelim=[is][\bfseries\textcolor{pred}]{/-}{-/}
}

\begin{document}
\title{Evaluating LLMs for One-Shot Patching of Real and Artificial Vulnerabilities}
%
%
\author{Aayush Garg}
\email{aayush.garg@list.lu}
\orcid{0000-0002-2507-8846}
\affiliation{%
  \institution{Luxembourg Institute of Science and Technology}
  \country{Luxembourg}
}

\author{Zanis Ali Khan}
\email{zanis-ali.khan@list.lu}
\orcid{0000-0002-3935-2148}
\affiliation{%
  \institution{Luxembourg Institute of Science and Technology}
  \country{Luxembourg}
}

\author{Renzo Degiovanni}
\email{renzo.degiovanni@list.lu}
\orcid{0000-0003-1611-3969}
\affiliation{%
  \institution{Luxembourg Institute of Science and Technology}
  \country{Luxembourg}
}

\author{Qiang Tang}
\email{qiang.tang@list.lu}
\orcid{0000-0002-6153-4255}
\affiliation{%
  \institution{Luxembourg Institute of Science and Technology}
  \country{Luxembourg}
}

%

%
\begin{abstract}
Automated vulnerability patching is crucial for software security, and recent advancements in Large Language Models (LLMs) present promising capabilities for automating this task. However, existing research has primarily assessed LLMs using publicly disclosed vulnerabilities, leaving their effectiveness on related artificial vulnerabilities largely unexplored. In this study, we empirically evaluate the patching effectiveness and complementarity of several prominent LLMs, such as OpenAI's GPT variants, LLaMA, DeepSeek, and Mistral models, using both real and artificial vulnerabilities. Our evaluation employs Proof-of-Vulnerability (PoV) test execution to concretely assess whether LLM-generated source code successfully patches vulnerabilities. Our results reveal that LLMs patch real vulnerabilities more effectively compared to artificial ones. Additionally, our analysis reveals significant variability across LLMs in terms of overlapping (multiple LLMs patching the same vulnerabilities) and complementarity (vulnerabilities patched exclusively by a single LLM), emphasizing the importance of selecting appropriate LLMs for effective vulnerability patching.

\end{abstract}
\maketitle              
\thispagestyle{preprint}
\section{Introduction}
Automated vulnerability patching has become increasingly significant in software security due to the escalating number and complexity of software vulnerabilities discovered annually. Recently, \textbf{Large Language Models \textit{(LLMs)}} including \textit{\textbf{OpenAI}}'s \textit{\textbf{GPT}} variants, \textit{\textbf{LLaMA}}, \textit{\textbf{DeepSeek}}, and \textit{\textbf{Mistral}}, have emerged as promising tools for automating vulnerability patching. These models have shown notable capability in generating syntactically and semantically coherent software patches~\cite{WuJPLD0BS23,abs-2501-03446}. However, the effectiveness of these models in patching vulnerabilities, particularly across diverse vulnerability types and their variants, remains understudied.

Previous studies evaluating LLM-based patching have primarily focused on known, i.e., publicly disclosed vulnerabilities sourced from public vulnerability databases~\cite{BuiSF22,abs-2408-13597,10.1145/2610384.2628055} or real-world software repositories~\cite{PearceTAKD23,abs-2501-07339,khan_multi-dataset_2025}. These evaluations typically consider syntactic correctness or employ similarity-based metrics such as \textit{CodeBLEU}~\cite{abs-2009-10297} and \textit{CrystalBLEU}~\cite{EghbaliP22}. While these metrics provide insights into patch quality, they do not directly indicate whether the generated patches \textbf{effectively eliminate} vulnerabilities~\cite{KulsumZXd24,abs-2405-01580,10.1145/3597503.3639182}.

Moreover, a significant limitation in existing literature is the scarcity of research investigating whether LLMs can generalize their vulnerability patching capabilities beyond these known vulnerabilities~\cite{wang2025vulnrepairevalexploitbasedevaluationframework}. \textit{Garg et al.}\cite{GargDPT24} recently augmented vulnerability dataset with their generated \textbf{artificial vulnerabilities}. We considered these artificial cases in our study to augment our evaluation and to test LLMs’ ability to patch both real and artificial vulnerabilities. This enabled us to perform an assessment of the generalizability and robustness of LLM-based vulnerability patching~\cite{abs-2312-15223}.

Our paper empirically investigates the effectiveness and complementarity of several prominent LLMs in automated vulnerability patching, specifically focusing on both real vulnerabilities and their corresponding artificial vulnerabilities. We structure our investigation around two research questions (RQs):

\begin{quote}
\begin{adjustwidth}{-0.5cm}{-0.5cm}
\textbf{RQ1: How effective are LLMs in patching real vulnerabilities vs. artificial vulnerabilities?}
\end{adjustwidth}
\end{quote}
RQ1 investigates the extent to which LLMs can successfully patch real vulnerabilities compared to their artificial counterparts, quantitatively comparing their effectiveness across these vulnerability categories.

\begin{quote}
\begin{adjustwidth}{-0.5cm}{-0.5cm}
\textbf{RQ2: How complementary and overlapping are LLMs in patching real vulnerabilities vs. artificial vulnerabilities?}
\end{adjustwidth}
\end{quote}
RQ2 explores how frequently multiple LLMs successfully patch the same vulnerabilities (\textbf{overlapping}), and how often vulnerabilities are successfully patched exclusively by a single LLM (\textbf{complementary}). By evaluating this distribution across real and artificial vulnerabilities, we assess whether different LLMs consistently patch the same vulnerabilities or contribute uniquely by patching distinct vulnerabilities.

To perform this evaluation, we employed \textbf{15} real vulnerabilities and their \textbf{41} artificial counterparts (vulnerabilities). We applied a uniform prompting strategy across all evaluated LLMs to generate patches. Evaluation of these patches was performed using PoV test executions, ensuring rigorous and concrete validation of the effectiveness of the generated patches.

Our contributions are summarized as follows:
\begin{enumerate}
    \item A systematic evaluation of the effectiveness of prominent LLMs in patching real and artificial vulnerabilities, measured using Proof-of-Vulnerability (PoV) test execution.
    \item An analysis of the extent to which LLM-generated patches overlap across models or complement each other, identifying cases where specific vulnerabilities are patched either by multiple LLMs or uniquely by individual LLMs.
\end{enumerate}

\section{Background}
In this section, we present the necessary background, including prior works on LLM-based vulnerability patching.
\subsection{Large Language Model \textit{(LLM)}-based Automated Vulnerability Patching}
Traditional vulnerability-patching methods in industry often rely on static analysis tools and manual expert fixes. Static analyzers can flag common weaknesses and even suggest remedial code patterns (e.g., adding input validations), though these suggestions are typically limited to well-known bug patterns~\cite{8952355}. General automated program repair tools have also been evaluated on security bugs, but their success is modest~\cite{10.1145/3597503.3639095}. For instance, Bui et al.~\cite{BuiPVMS24} found that state-of-the-art automated vulnerability patching tools could produce testable patches for only about \textbf{20\%} of real-world vulnerabilities, and under half of those patches ($\sim$44\%) actually fixed the issue without breaking functionality. This amounts to less than 10\% of vulnerabilities being fully remedied by traditional APR techniques. In addition, domain-specific patch generation has seen some success in narrow contexts, e.g., automatic insertion of security checks to fix Android component-hijacking flaws, effectively patching such vulnerabilities in multiple real-world apps~\cite{ZhangY14}. These targeted solutions, however, are limited to specific vulnerability types.

LLMs have demonstrated significant potential in program repair, particularly in automated vulnerability patching~\cite{WuJPLD0BS23,abs-2501-03446,10.1145/3524459.3527351}. Given a vulnerable code snippet, these models generate a patched version that aims to eliminate security flaws while preserving functionality. Prior research has evaluated LLMs' ability to patch vulnerabilities based on datasets of publicly disclosed vulnerabilities~\cite{PearceTAKD23,abs-2408-13597}. However, LLM-generated patches often vary in effectiveness, requiring validation through test cases to confirm whether the vulnerability has been successfully addressed~\cite{KulsumZXd24,abs-2405-01580,10.5555/3766078.3766309}.

Despite promising results, LLM-based patching faces challenges. First, LLMs may \textbf{overfit to known vulnerabilities}, producing patches that resemble existing solutions rather than addressing the underlying security flaw~\cite{abs-2501-03446}. Second, \textbf{syntactic correctness does not guarantee functional correctness}, making execution-based evaluation crucial~\cite{DibiaFBPLA23,OjdanicGKDPT23,10173980}. Lastly, the generalization of LLMs to variations of real vulnerabilities remains an open question~\cite{abs-2409-10756}, which this study seeks to investigate.

\subsection{Artificial Vulnerabilities}
Artificial vulnerabilities are intentionally introduced flaws in a system, such as source code, designed to mimic real security weaknesses~\cite{GargDPT24,10.1145/3135932.3135941,abs-2303-04247}. These controlled weaknesses allow researchers and developers to \textbf{assess and refine detection and mitigation} strategies in a reproducible environment~\cite{GargDJCPT22,YE2021110825,abs-2012-11701}.

Several methodologies exist for \textbf{creating} artificial vulnerabilities. \textbf{Code injection} introduces malicious code to simulate attacks like SQL injection or cross-site scripting~\textit{(XSS)}, helping assess defensive measures~\cite{HalfondVO06}. Fault injection deliberately introduces errors to test system resilience and error-handling capabilities~\cite{HsuehTI97}. \textbf{Mutation testing}, on the other hand, modifies source code systematically to replicate common programming mistakes, evaluating the effectiveness of testing strategies~\cite{JiaH11,10.1145/2635868.2635929,GargODCPT23}. For example, the \textit{LAVA-M}~\cite{7546498} dataset systematically injects synthetic memory-safety bugs into C programs to create ground-truth vulnerability cases. These techniques collectively support the development of more secure and robust software systems.

\subsection{Evaluation of LLM-Based Patching}
Existing studies primarily evaluate LLM-based code generation using \textbf{code similarity metrics} such as \textit{\textbf{CodeBLEU}}~\cite{abs-2009-10297} and \textit{\textbf{CrystalBLEU}}~\cite{EghbaliP22}, which assess the syntactic and semantic resemblance between the generated code and the expected. However, these metrics may not accurately reflect the security efficacy of patches, as they do not account for the actual elimination of vulnerabilities~\cite{NasrabadiPRRE23}.

To ensure that patches genuinely address security flaws, \textbf{test execution validation} is necessary~\cite{DibiaFBPLA23,10.1007/s10115-025-02383-9,wang2025vulnrepairevalexploitbasedevaluationframework}. This approach involves running \textbf{Proof-of-Vulnerability (\textit{PoV})} test cases against the patched code to verify that the specific vulnerabilities have been effectively mitigated. Such execution-based evaluations provide a more reliable assessment of the patch's effectiveness in real-world scenarios~\cite{LiuXW023,10.1145/3704997}. \textit{Wang et al.}~\cite{wang2025vulnrepairevalexploitbasedevaluationframework} also demonstrated that without rigorous exploit-driven tests, an LLM-generated patch may give a \textbf{false sense of safety} where it can appear to fix the bug while the underlying vulnerability remains exploitable. Hence, in this study, we prioritize the \textit{PoV} test results over similarity-based evaluations to assess the effectiveness of LLM-generated patches.

\section{Methodology}
This section describes our approach for evaluating LLM patching effectiveness, detailing the dataset, models, prompting strategy, and evaluation metrics.

\subsection{Dataset and Vulnerabilities}
Firstly, we utilized the \textit{\textbf{Vul4J}} dataset~\cite{BuiSF22}, a carefully-curated benchmark of \textbf{reproducible} Java vulnerabilities drawn from open-source projects and covering distinct Common Weakness Enumeration (CWE) classes.  For every entry Vul4J provides (i) the vulnerable revision (${V_\text{vul}}$), (ii) the corresponding fixed revision (${V_\text{fix}}$), (iii) the minimal human-authored patch, and crucially (iv) one or more \textbf{Proof-of-Vulnerability \textit{(PoV)}} JUnit test cases. A PoV test is an executable exploit oracle: it exercises the buggy code so that the test \textbf{fails} on ${V_\text{vul}}$ (e.g., by throwing an exception, timing out, or violating an assertion) and \textbf{passes} on ${V_\text{fix}}$. This PoV-test fail-to-pass flip allows automated repair systems to verify, with a single test run, whether a candidate patch truly eliminates the vulnerability. When no suitable test existed in the original project, \textit{Bui et al.}~\cite{BuiSF22} manually wrote one by following the exploit steps in the CVE report, thereby guaranteeing that every vulnerability in the dataset can be triggered, reproduced, and validated in a \textit{Maven/Gradle} build.

\begin{figure*}[t!]
\centering
\begin{subfigure}[t]{0.32\textwidth}
    \begin{lstlisting}[language=PrettyJava]
private static void decompress
 (final InputStream in, final byte[] out)
 throws IOException {
  int position = 0;
  final int total = out.length;
  while (position < total) {
   final int n = in.read();


   
   if (n > 128) {
    final int value = in.read();
    for (int i = 0; i < (n & 0x7f); i++) {
     out[position++] = (byte) value; }
   } else {
    for (int i = 0; i < n; i++) {
    out[position++] = (byte) in.read();}
   }
  }
 }
\end{lstlisting}
    \caption{Real Vulnerability (CVE-2018-17201)}
    \label{fig_example_1_a}
  \end{subfigure}
\hfill
\begin{subfigure}[t]{0.32\textwidth}
    \begin{lstlisting}[language=PrettyJava]
private static void decompress
 (final InputStream in, final byte[] out)
 throws IOException {
  int position = 0;
  final int total = out.length;
  while (position < total) {
   final int n = in.read();
   /+if (n < 0) {
    throw new ImageReadException("Error decompressing RGBE file"); }+/
   if (n > 128) {
    final int value = in.read();
    for (int i = 0; i < (n & 0x7f); i++) {
     out[position++] = (byte) value; }
   } else {
    for (int i = 0; i < n; i++) {
    out[position++] = (byte) in.read();}
   }
  }
 }
\end{lstlisting}
    \caption{Patched source code}
    \label{fig_example_1_b}
  \end{subfigure}
\hfill
\begin{subfigure}[t]{0.32\textwidth}
\begin{lstlisting}[language=PrettyJava]
private static void decompress
 (final InputStream in, final byte[] out)
 throws IOException {
  int position = 0;
  final int total = out.length;
  while (position < total) {
   final int n = in.read();
   if (n /-==-/ 0) { // `<' modified to `=='
    throw new ImageReadException("Error decompressing RGBE file"); }
   if (n > 128) {
    final int value = in.read();
    for (int i = 0; i < (n & 0x7f); i++) {
     out[position++] = (byte) value; }
   } else {
    for (int i = 0; i < n; i++) {
    out[position++] = (byte) in.read();}
   }
  }
 }
\end{lstlisting}
    \caption{Artificial Vulnerability}
    \label{fig_example_1_c}
  \end{subfigure}
  \vspace{-1em}
\caption{Vulnerability CVE-2018-17201 (Fig. \ref{fig_example_1_a}) causes an infinite loop that can make the program hang, enabling a Denial-of-Service (DoS) attack. This issue is addressed by introducing a conditional exception using an ``\textit{if}'' statement (Fig. \ref{fig_example_1_b}). However, the code-modification (Fig. \ref{fig_example_1_c}) alters the ``\textit{if}'' condition, effectively nullifying the fix and reintroduces the vulnerability.}
\label{fig_example_1}
\end{figure*}

\begin{figure*}[t!]
\centering
\begin{subfigure}[t]{0.32\textwidth}
    \begin{lstlisting}[language=PrettyJava]
void addPathParam(String name, String value, boolean encoded) {
 if (relativeUrl == null) {
  throw new AssertionError(); }





 /-relativeUrl = relativeUrl.replace("{" + name + "}", canonicalizeForPath(value, encoded));-/
 



}
\end{lstlisting}
    \caption{Real Vulnerability (CVE-2018-1000850)}
    \label{fig_example_2_a}
  \end{subfigure}
\hfill
\begin{subfigure}[t]{0.32\textwidth}
    \begin{lstlisting}[language=PrettyJava]
void addPathParam(String name, String value, boolean encoded) {
 if (relativeUrl == null) {
  throw new AssertionError(); }
 /+String replacement = canonicalizeForPath(value, encoded);
 String newRelativeUrl = relativeUrl.replace("{" + name + "}", replacement);
 if (PATH_TRAVERSAL
  .matcher(newRelativeUrl)
  .matches()) {
   throw new IllegalArgumentException(
   "@Path parameters shouldn't perform path traversal (`.' or `..'): " + value); }
 relativeUrl = newRelativeUrl;+/
}
\end{lstlisting}
    \caption{Patched source code}
    \label{fig_example_2_b}
  \end{subfigure}
\hfill
\begin{subfigure}[t]{0.32\textwidth}
\begin{lstlisting}[language=PrettyJava]
void addPathParam(String name, String value, boolean encoded) {
 if (relativeUrl == null) {
  throw new AssertionError(); }
 String replacement = canonicalizeForPath(value, encoded);
 String newRelativeUrl = relativeUrl.replace("{" + name + "}", replacement);
 if (PATH_TRAVERSAL
  .matcher(/-name-/)//passed argument changed
  .matches()) {
   throw new IllegalArgumentException(
   "@Path parameters shouldn't perform path traversal (`.' or '..'): " + value); }
 relativeUrl = newRelativeUrl;
}
\end{lstlisting}
    \caption{Artificial Vulnerability}
    \label{fig_example_2_c}
  \end{subfigure}
  \vspace{-1em}
\caption{Vulnerability CVE-2018-1000850 (Fig. \ref{fig_example_2_a}) enables path traversal and access to restricted directories. The patch adds a conditional check for `\texttt{.}' or `\texttt{..}' in ``\textit{newRelativeUrl}'' (Fig. \ref{fig_example_2_b}). The code-modification (Fig. \ref{fig_example_2_c}) replaces ``\textit{newRelativeUrl}'' with ``\textit{name}'', nullifying the fix and reintroducing the vulnerability.}
\label{fig_example_2}
\vspace{-1em}
\end{figure*}

Secondly, we augmented our evaluation with \textbf{artificially} generated vulnerabilities to broaden our test criteria. By including the artificial cases, we wanted to test LLMs’ ability to patch both real and, artificial yet representative, vulnerabilities. Additionally, we wanted to observe whether LLMs handle them differently. To do so, we incorporated artificial vulnerabilities derived from the work of \textit{Garg et al.}\cite{GargDPT24}. \textit{Garg et al.} used \textit{\textbf{CodeBERT}}\cite{FengGTDFGS0LJZ20} to generate thousands of candidate artificial vulnerabilities and validated them with Vul4J’s PoV tests. Only about 4\% of the candidates reproduced the \textbf{same} PoV failures (including identical exception messages) observed for real Vul4J vulnerabilities. Those validated cases cover \textbf{15} real CVEs, i.e., roughly 55\% of Vul4J's vulnerabilities.

To illustrate, \textbf{Figures~\ref{fig_example_1} and~\ref{fig_example_2}} provide motivating examples of artificial and real vulnerabilities, originally presented by \textit{Garg et al.}\cite{GargDPT24}. Figure~\ref{fig_example_1} shows the code-modification that nullifies a patch for a high-severity (\textbf{CVSS 7.5}) infinite loop vulnerability (\textit{\textbf{CVE-2018-17201}}~\cite{CVE-2018-17201}) by altering a conditional check, effectively reintroducing the real vulnerability. Similarly, Figure~\ref{fig_example_2} demonstrates the code-modification that reintroduces a directory traversal vulnerability (\textit{\textbf{CVE-2018-1000850}}~\cite{CVE-2018-1000850}) by modifying an argument in a validation function. 

\begin{table*}[tp]
\centering
\caption{Overview of real and artificial vulnerabilities}
\vspace{-1em}
\label{tab:original_cwe_artificial}
\resizebox{\textwidth}{!}{
\begin{tabular}{l|l|l|c|c|c}
\toprule
\makecell{\textbf{Real}\\\textbf{Vulnerability}\\\textbf{(CVE ID)}} & 
\makecell{\textbf{Vulnerability}\\\textbf{Class}\\\textbf{(CWE ID)}} &
\makecell{\textbf{CWE Description}\\\textbf{(Vulnerability Cause)}} &
\makecell{\textbf{Severity}\\\textbf{(0 -- 10)}} &
\makecell{\textbf{Failed}\\\textbf{Tests}\\\textbf{(PoV)}} &
\makecell{\textbf{Artificial}\\\textbf{Vulnerabilities}\\\textbf{(\#)}} \\
\midrule
APACHE-COMMONS-001 & CWE-noinfo & \makecell[l]{No information provided by NIST} & NA & 1 & 1 \\ \hline
CVE-2013-5960 & CWE-310 & \makecell[l]{Cryptographic Issues} & 5.8 & 15 & 1 \\ \hline
CVE-2014-4172 & CWE-74 & \makecell[l]{Improper neutralization of special elements in output\\used by a downstream component (`Injection')} & 9.8 & 1 & 10 \\ \hline
CVE-2016-10006 & CWE-79 & \makecell[l]{Improper neutralization of input during web page\\generation (`Cross-site scripting')} & 6.1 & 1 & 1 \\ \hline
CVE-2016-2162 & CWE-79 & \makecell[l]{Improper neutralization of input during web page\\generation (`Cross-site scripting')} & 6.1 & 1 & 1 \\ \hline
CVE-2016-6802 & CWE-284 & \makecell[l]{Improper access control} & 7.5 & 3 & 1 \\ \hline
CVE-2017-5662 & CWE-611 & \makecell[l]{Improper restriction of xml external entity reference} & 7.3 & 1 & 6 \\ \hline
CVE-2018-1000089 & CWE-532 & \makecell[l]{Insertion of sensitive information into log file} & 7.4 & 1 & 7 \\ \hline
CVE-2018-1000531 & CWE-20 & \makecell[l]{Improper input validation} & 7.5 & 1 & 1 \\ \hline
CVE-2018-1000850 & CWE-22 & \makecell[l]{Improper limitation of a pathname to a restricted\\directory (`Path traversal')} & 7.5 & 3 & 1 \\ \hline
CVE-2018-1000854 & CWE-74 & \makecell[l]{Improper neutralization of special elements in output\\used by a downstream component (`Injection')} & 9.8 & 1 & 1 \\ \hline
CVE-2018-11771 & CWE-835 & \makecell[l]{Loop with unreachable exit condition (`Infinite loop')} & 5.5 & 2 & 2 \\ \hline
CVE-2018-17201 & CWE-835 & \makecell[l]{Loop with unreachable exit condition (`Infinite loop')} & 7.5 & 1 & 2 \\ \hline
CVE-2019-12402 & CWE-835 & \makecell[l]{Loop with unreachable exit condition (`Infinite loop')} & 7.5 & 1 & 1 \\ \hline
HTTPCLIENT-1803 & CWE-noinfo & \makecell[l]{No information provided by NIST} & NA & 1 & 5 \\
\midrule
\multicolumn{3}{c}{\textbf{Total Real Vulnerabilities}} & \multicolumn{3}{c}{\textbf{15}} \\
\multicolumn{3}{c}{\textbf{Total Artificial Vulnerabilities}} & \multicolumn{3}{c}{\textbf{41}} \\
\bottomrule
\end{tabular}
}
\end{table*}

Please \textbf{note} that we only considered these 4\% artificial vulnerabilities from \textit{Garg et al.}'s dataset that exhibit the \textbf{same} failure patterns as the real vulnerabilities. \textit{Vul4J v1.1} currently contains 27 reproducible Java vulnerabilities; the 15 we study are those for which validated artificial vulnerabilities exist. This resulted in our dataset of \textbf{15} real vulnerabilities (from \textit{\textbf{Vul4J}}) and their corresponding \textbf{41} artificial vulnerabilities (from \textit{Garg et al.}), for which we present the details (i.e., vulnerability's CVE ID, CWE details, severity, failed PoV tests, and the number of corresponding artificial cases) in \textbf{Table~\ref{tab:original_cwe_artificial}}.

\subsection{LLMs Evaluated} 
We evaluate the patching capabilities of \textbf{14} large language models~\textit{(LLMs)} spanning different architectures and sizes. These include models from \textit{OpenAI}’s GPT series, \textit{Meta}’s \textit{LLaMA} models, \textit{DeepSeek}, and \textit{Mistral}’s instruction-tuned variants. Below are the LLMs that we assess in our study.
\begin{enumerate}
    \item \textbf{DeepSeekR1 Qwen 32B}
    \item \textbf{GPT3.5 Turbo}
    \item \textbf{GPT3.5 Turbo 1106}
    \item \textbf{GPT3.5 Turbo Instruct}
    \item \textbf{GPT4}
    \item \textbf{GPT4 0613}
    \item \textbf{GPT4 Turbo}
    \item \textbf{GPT4o}
    \item \textbf{GPT4o Mini}
    \item \textbf{LLaMA 3.1 70B Instruct}
    \item \textbf{LLaMA 3.3 70B Instruct}
    \item \textbf{Mistral 7B v0.2 Instruct}
    \item \textbf{Mistral 7B v0.3 Instruct}
    \item \textbf{Mistral 8×7B v0.1 Instruct}
\end{enumerate}
These \textbf{14} LLMs were selected to cover a range of training paradigms, instruction-tuned capabilities, and underlying architectures, allowing for a comprehensive analysis of their effectiveness in vulnerability patching.

\subsection{Prompting Strategy}

To ensure a consistent evaluation across all models, we employed a standardized prompt designed to elicit vulnerability patches while minimizing extraneous modifications:
\begin{quote}
\begin{adjustwidth}{-0.5cm}{-0.5cm}
\texttt{\textit{"You are a security expert who is good at static program analysis. The following SOURCE CODE, written in Java, contains a vulnerability. Please write the source code to fix this vulnerability. Do not make any other changes to the source code, and just reply with the final patched Java function.}}

\texttt{\textit{SOURCE CODE:}} \{source code\}"
\end{adjustwidth}
\end{quote}

We \textbf{deliberately avoid advanced prompting/decoding} such as \textit{self-consistency}~\cite{wang2023selfconsistencyimproveschainthought} and \textit{iterative self-feedback} (\textit{Reflexion}~\cite{10.5555/3666122.3666499}, \textit{Self-Refine}~\cite{10.5555/3666122.3668141}), and evaluate each model in a \textbf{one-shot, prompt-only} setting to ensure comparability across models. Each LLM received only the vulnerable function’s Java source code. No surrounding files, build scripts, or PoV tests were supplied. This prompt explicitly instructs the LLM to focus solely on patching the vulnerability without altering unrelated code segments. Each LLM was prompted separately for every vulnerability, ensuring that the responses were generated independently. The collected outputs were then assessed for their patching success across corresponding vulnerabilities.

\subsection{Evaluation Process}

The patches generated by the LLMs were compiled and executed against the Proof-of-Vulnerability (PoV) test cases. For every vulnerability, we programmatically replaced the original function with the model‑generated version (patch), rebuilt the project, and executed the PoV suite. A patch was considered \textbf{successful} if it successfully compiled and passed all associated PoV tests. If a model’s output resulted in compilation errors or failed the PoV test cases, it was considered unsuccessful.

\subsection{Evaluation Metrics}
To assess the patching capabilities of LLMs, we employ the following key metrics:

\begin{enumerate}
    \item \textbf{Patching Success}: The percentage of vulnerabilities successfully patched by each model, measured separately for real and artificial vulnerabilities, enabling a direct comparison between the two categories.
    \item \textbf{Overlapping Patching}: The proportion of vulnerabilities that were patched by multiple models, indicating common success cases.
    \item \textbf{Complementary Patching}: The proportion of vulnerabilities patched exclusively by a single model, highlighting the complementary contributions to patching.
\end{enumerate}

These metrics directly address our research questions, providing insights into the effectiveness and complementarity of LLMs in patching both real and artificial vulnerabilities.

\begin{table*}[tp]
\centering
\caption{Overview of LLMs in patching real and artificial vulnerabilities}
\label{tab:patch_matrix}
\resizebox{\textwidth}{!}{
\begin{tabular}{l|c|c|c|c|c|c|c|c|c|c|c|c|c|c}
\toprule
\textbf{Vuln.} 
& \makecell{\textbf{DeepSeek} \textbf{R1}\\\textbf{Qwen} \textbf{32B}} 
& \makecell{\textbf{GPT} \textbf{3.5}\\\textbf{Turbo}} 
& \makecell{\textbf{GPT} \textbf{3.5}\\\textbf{Turbo} \textbf{1106}} 
& \makecell{\textbf{GPT} \textbf{3.5}\\\textbf{Turbo} \textbf{Instruct}} 
& \makecell{\textbf{GPT} \textbf{4}} 
& \makecell{\textbf{GPT} \textbf{4}\\\textbf{0613}} 
& \makecell{\textbf{GPT} \textbf{4}\\\textbf{Turbo}} 
& \makecell{\textbf{GPT} \textbf{4o}} 
& \makecell{\textbf{GPT} \textbf{4o}\\\textbf{Mini}} 
& \makecell{\textbf{LLaMA} \textbf{3.1}\\\textbf{70B} \textbf{Instruct}} 
& \makecell{\textbf{LLaMA} \textbf{3.3}\\\textbf{70B} \textbf{Instruct}} 
& \makecell{\textbf{Mistral} \textbf{7B}\\\textbf{v0.2} \textbf{Instruct}} 
& \makecell{\textbf{Mistral} \textbf{7B}\\\textbf{v0.3} \textbf{Instruct}} 
& \makecell{\textbf{Mistral} \textbf{8x7B}\\\textbf{v0.1} \textbf{Instruct}} \\

\midrule
\multicolumn{15}{c}{\textbf{APACHE-COMMONS-001}} \\
real &  &  &  &  &  &  &  &  &  &  &  &  &  &  \\
artificial &  &  &  &  &  &  &  &  &  &  &  &  &  &  \\

\midrule
\multicolumn{15}{c}{\textbf{CVE-2013-5960}} \\
real & \checkmark & \checkmark & \checkmark & \checkmark & \checkmark & \checkmark & \checkmark & \checkmark & \checkmark & \checkmark & \checkmark & \checkmark & \checkmark & \checkmark \\
artificial & \checkmark & \checkmark & \checkmark & \checkmark & \checkmark & \checkmark & \checkmark & \checkmark & \checkmark & \checkmark & \checkmark & \checkmark & \checkmark & \checkmark \\

\midrule
\multicolumn{15}{c}{\textbf{CVE-2014-4172}} \\
real &  &  &  &  &  &  &  &  &  &  &  &  &  &  \\
artificial 01 &  &  &  &  &  &  &  &  &  &  &  &  &  &  \\
artificial 02 &  &  &  &  &  &  &  &  &  &  &  &  &  &  \\
artificial 03 &  &  &  &  &  &  &  &  &  &  &  &  &  &  \\
artificial 04 &  &  &  &  &  &  &  &  &  &  &  &  &  &  \\
artificial 05 &  &  &  &  &  &  &  &  &  &  &  &  &  &  \\
artificial 06 &  &  &  &  &  &  &  &  &  &  &  &  &  &  \\
artificial 07 &  &  &  &  &  &  &  &  &  &  &  &  &  &  \\
artificial 08 &  &  &  &  &  &  &  &  &  &  &  &  &  &  \\
artificial 09 &  &  &  &  &  &  &  &  &  &  &  &  &  &  \\
artificial 10 &  &  &  &  &  &  &  &  &  &  &  &  &  &  \\

\midrule
\multicolumn{15}{c}{\textbf{CVE-2016-10006}} \\
real &  &  &  &  &  &  &  &  &  &  &  &  &  &  \\
artificial &  &  &  &  &  & \checkmark &  & \checkmark &  &  &  &  & \checkmark & \checkmark \\

\midrule
\multicolumn{15}{c}{\textbf{CVE-2016-2162}} \\
real &  &  &  &  &  &  &  &  &  &  &  &  &  &  \\
artificial & \checkmark &  &  &  &  &  &  &  &  &  &  &  & \checkmark &  \\

\midrule
\multicolumn{15}{c}{\textbf{CVE-2016-6802}} \\
real &  &  &  &  & \checkmark &  & \checkmark &  &  &  &  & \checkmark &  & \checkmark \\
artificial &  &  &  &  & \checkmark &  & \checkmark &  &  &  &  &  &  & \checkmark \\

\midrule
\multicolumn{15}{c}{\textbf{CVE-2017-5662}} \\
real &  &  &  &  &  &  &  &  &  &  &  &  &  &  \\
artificial 01 &  &  &  &  &  &  &  &  &  &  &  &  &  &  \\
artificial 02 &  &  &  &  &  &  &  &  &  &  &  &  &  &  \\
artificial 03 &  &  &  &  &  &  &  &  &  &  &  &  &  &  \\
artificial 04 &  &  &  &  &  &  &  &  &  &  &  &  &  &  \\
artificial 05 &  &  &  &  &  &  &  &  &  &  &  &  &  &  \\
artificial 06 &  &  &  &  &  &  &  &  &  &  &  &  &  &  \\

\midrule
\multicolumn{15}{c}{\textbf{CVE-2018-1000089}} \\
real &  &  &  &  &  &  &  &  &  &  &  &  &  &  \\
artificial 01 &  &  &  &  &  &  &  &  &  &  &  &  &  &  \\
artificial 02 &  &  &  &  &  &  &  &  &  &  &  &  &  &  \\
artificial 03 &  &  &  &  &  &  &  &  &  &  &  &  &  &  \\
artificial 04 &  &  &  &  &  &  &  &  &  &  &  &  &  &  \\
artificial 05 &  &  &  &  &  &  &  &  &  &  &  &  &  &  \\
artificial 06 &  &  &  &  &  &  &  &  &  &  &  &  &  &  \\
artificial 07 &  &  &  &  &  &  &  &  &  &  &  &  &  &  \\

\midrule
\multicolumn{15}{c}{\textbf{CVE-2018-1000531}} \\
real & \checkmark &  &  & \checkmark & \checkmark & \checkmark &  & \checkmark &  & \checkmark & \checkmark & \checkmark & \checkmark & \checkmark \\
artificial &  &  &  &  &  &  &  &  &  &  &  &  &  &  \\

\midrule
\multicolumn{15}{c}{\textbf{CVE-2018-1000850}} \\
real & \checkmark &  & \checkmark & \checkmark & \checkmark & \checkmark & \checkmark & \checkmark & \checkmark & \checkmark & \checkmark & \checkmark & \checkmark & \checkmark \\
artificial &  &  &  &  &  &  &  &  &  &  &  &  &  &  \\

\midrule
\multicolumn{15}{c}{\textbf{CVE-2018-1000854}} \\
real & \checkmark & \checkmark & \checkmark & \checkmark & \checkmark & \checkmark & \checkmark & \checkmark & \checkmark & \checkmark & \checkmark & \checkmark & \checkmark & \checkmark \\
artificial &  &  &  &  &  &  &  &  &  &  &  &  &  &  \\

\midrule
\multicolumn{15}{c}{\textbf{CVE-2018-11771}} \\
real &  &  &  &  &  &  &  &  &  &  &  &  & \checkmark &  \\
artificial 1 &  &  &  &  &  &  &  &  &  &  &  &  &  &  \\
artificial 2 &  &  &  &  &  &  &  &  &  &  &  &  &  &  \\

\midrule
\multicolumn{15}{c}{\textbf{CVE-2018-17201}} \\
real & \checkmark & \checkmark &  & \checkmark & \checkmark &  & \checkmark & \checkmark & \checkmark & \checkmark & \checkmark & \checkmark &  & \checkmark \\
artificial 1 & \checkmark & \checkmark & \checkmark &  &  & \checkmark &  &  &  &  &  & \checkmark &  & \checkmark \\
artificial 2 &  &  &  &  &  &  &  &  &  &  &  &  &  &  \\

\midrule
\multicolumn{15}{c}{\textbf{CVE-2019-12402}} \\
real &  &  &  &  &  &  &  &  &  &  &  &  &  &  \\
artificial &  &  &  &  &  &  &  &  &  &  &  &  &  &  \\

\midrule
\multicolumn{15}{c}{\textbf{HTTPCLIENT-1803}} \\
real & \checkmark &  & \checkmark &  &  &  &  &  &  &  &  &  &  & \checkmark \\
artificial 1 & \checkmark &  &  &  & \checkmark &  &  &  & \checkmark &  &  &  &  & \checkmark \\
artificial 2 & \checkmark &  &  &  &  &  & \checkmark &  &  & \checkmark &  &  &  & \checkmark \\
artificial 3 & \checkmark &  &  &  &  &  & \checkmark &  &  & \checkmark &  &  &  &  \\
artificial 4 & \checkmark &  &  &  &  &  &  &  &  &  &  &  &  & \checkmark \\
artificial 5 & \checkmark &  &  &  &  &  &  &  &  &  &  &  &  &  \\

\bottomrule
\end{tabular}
}
\end{table*}

\begin{figure*}[tp]
    \centering
    \includegraphics[width=\textwidth]{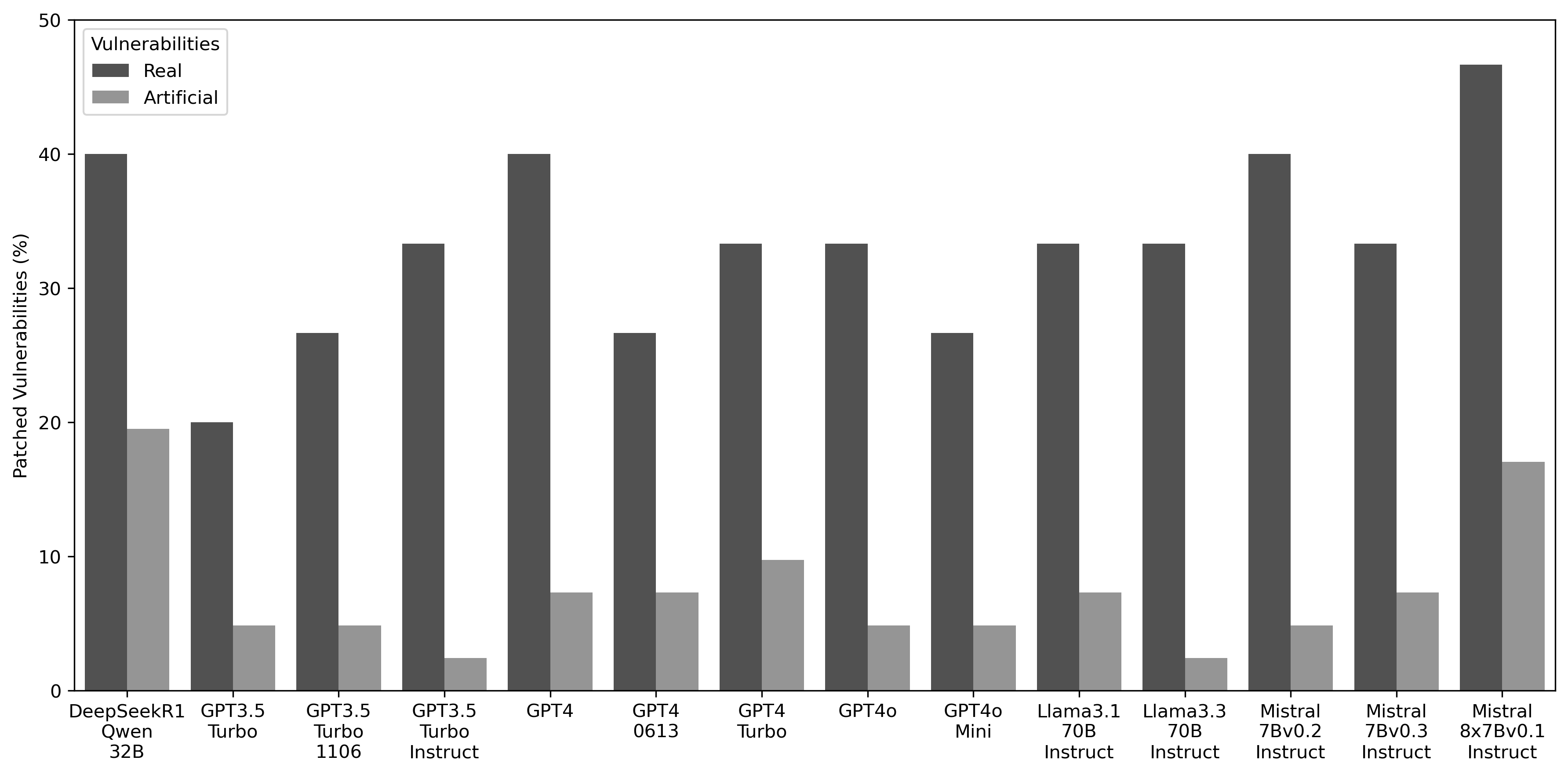}
    \vspace{-2.5em}
    \caption{Patching effectiveness of LLMs for real and artificial vulnerabilities.}
    \label{fig:plot_rq1}
\vspace{-1em}
\end{figure*}

\section{Results}

\textbf{Table~\ref{tab:patch_matrix}} provides an overview of different LLMs patching real and artificial vulnerabilities. Each section is grouped by CVE ID, covering both real and artificial vulnerabilities. The first column specifies the vulnerability type, distinguishing between real and artificial cases. The remaining columns represent different LLMs, where a checkmark (\checkmark) indicates successful patching of the vulnerability by that model, while an empty cell denotes failure to patch, either due to a compilation error or failed test cases (Proof-of-Vulnerability). Moreover, the table enables a structured representation of how different models overlap or complement each other in their patching capabilities. The table reveals \textbf{three} trends: 
\begin{enumerate}
    \item \textbf{Real vs. artificial.} Check-marks are still skewed toward the \textbf{real} CVEs, i.e., more than half (8/15, $\sim$53\%) of the real vulnerabilities receive at least one patch, whereas fewer than one-quarter (10/41, $\sim$24\%) of the artificial vulnerabilities are patched. Entire blocks of artificial cases such as all \textbf{ten} variations of \textit{CVE-2014-4172} and \textbf{seven} of \textit{CVE-2018-1000089}, remain completely blank in the table, conveying LLMs' \textbf{failure to patch} such cases.
    \item \textbf{LLM spread.} As per our results, no single vendor dominated. The best overall scores were achieved \textit{DeepSeekR1 Qwen 32B} and \textit{Mistral 8x7Bv0.1 Instruct} with 14 successful patches each, followed by the GPT4/Turbo with 9 patches, and LLaMA3.1 70B Instruct and Mistral 7B variants with 8 patches, followed by the rest.
   \item \textbf{Overlap.}  When a real vulnerability was patched it was usually patched by \textbf{several} models, e.g., \textit{CVE-2018-1000531}, \textit{CVE-2018-1000850}, \textit{CVE-2018-1000854} were patched by almost every LLM, while the few artificial vulnerabilities that were patched were usually fixed by only one model (i.e., model overlap is very low for these cases).
\end{enumerate}

\paragraph{\textbf{Illustrative Examples.}}
We further inspected one success and one failure, for which, we provide the details below:
\begin{enumerate}
    \item \textbf{CVE-2013-5960 - patching-success.} \textit{OWASP ESAPI} encrypted data with the legacy mode \texttt{AES/CBC/PKCS5Padding} and accepted message-authentication tags shorter than 128 bits. An attacker could therefore flip cipher-text bits and tamper with the decrypted message. The official patch makes \textbf{two} tiny edits: (i) change the constant that selects the cipher to \texttt{AES/GCM/NoPadding}, an authenticated-encryption mode; and (ii) hard-code the tag length check to 128 bits. Every LLM considered in our study reproduced those \textbf{exact} two lines.  After applying the patch, the \textit{Proof-of-Vulnerability (PoV)} test, which tries to modify a valid cipher-text and then decrypt it, \textbf{fails as expected}, confirming the tampering attack is blocked.
    \item \textbf{CVE-2017-5662 - patching-failure.} Apache Batik’s XML parser resolved external entities by default, so a crafted SVG file could read and embed the contents of an arbitrary local file (e.g.\ \texttt{/etc/passwd}), an \textit{XML external entity injection (XXE)} vulnerability. A \textbf{complete} repair needs two coordinated changes: (i) in code, disable external-entity resolution (e.g., call \textit{setFeature(FEATURE\_SECURE\_PROCESSING, true)}); and (ii) in the build file, upgrade Batik to version $\ge$\,1.9, because older versions silently ignore that security flag in some pipelines. None of the LLMs produced both edits in the same patch where most toggled the parser flag but left the dependency unchanged. This likely reflects the fact that the required library upgrade resides outside the provided code snippet. When executed, the PoV tests failed (in contrast to passing-PoV-tests for a correct patch) with the malicious SVG still succeeding in printing the local file, hence indicating a failed patch.
\end{enumerate}

\begin{table*}[tp]
\centering
\caption{Overview of patched vulnerabilities by LLMs}
\vspace{-1em}
\label{tab:patching_rates}
\resizebox{\textwidth}{!}{
\begin{tabular}{l|c|c|c}
\toprule
\makecell{\textbf{Large Language Model}\\\textit{(LLM)}} &
\makecell{\textbf{Overall Vulnerabilities Patched}\\\textit{Count \# (Percentage \%)}} &
\makecell{\textbf{Real Vulnerabilities Patched}\\\textit{Count \# (Percentage \%)}} &
\makecell{\textbf{Artificial Vulnerabilities Patched}\\\textit{Count \# (Percentage \%)}} \\
\midrule
     DeepSeekR1 Qwen 32B &   14 (25.00\%) &   6 (40.00\%) &   8 (19.51\%) \\
           GPT3.5 Turbo &     5 (8.93\%) &   3 (20.00\%) &    2 (4.88\%) \\
       GPT3.5 Turbo 1106 &    6 (10.71\%) &   4 (26.67\%) &    2 (4.88\%) \\
   GPT3.5 Turbo Instruct &    6 (10.71\%) &   5 (33.33\%) &    1 (2.44\%) \\
                 GPT4 &    9 (16.07\%) &   6 (40.00\%) &    3 (7.32\%) \\
             GPT4 0613 &    7 (12.50\%) &   4 (26.67\%) &    3 (7.32\%) \\
            GPT4 Turbo &    9 (16.07\%) &   5 (33.33\%) &    4 (9.76\%) \\
                GPT4o &    7 (12.50\%) &   5 (33.33\%) &    2 (4.88\%) \\
            GPT4o Mini &    6 (10.71\%) &   4 (26.67\%) &    2 (4.88\%) \\
    LLaMA3.1 70B Instruct &    8 (14.29\%) &   5 (33.33\%) &    3 (7.32\%) \\
    LLaMA3.3 70B Instruct &    6 (10.71\%) &   5 (33.33\%) &    1 (2.44\%) \\
  Mistral 7Bv0.2 Instruct &    8 (14.29\%) &   6 (40.00\%) &    2 (4.88\%) \\
  Mistral 7Bv0.3 Instruct &    8 (14.29\%) &   5 (33.33\%) &    3 (7.32\%) \\
Mistral 8x7Bv0.1 Instruct &   14 (25.00\%) &   7 (46.67\%) &   7 (17.07\%) \\
\bottomrule
\end{tabular}
}
\end{table*}

These examples show that LLMs handle single-location fixes quite well. However, their success rate drops when a vulnerability demands several interdependent edits scattered across the codebase.

\begin{quote}
\begin{adjustwidth}{-0.5cm}{-0.5cm}
\textbf{RQ1: How effective are LLMs in patching real vulnerabilities vs. artificial vulnerabilities?}
\end{adjustwidth}
\end{quote}
To assess how well LLMs patch real vulnerabilities compared to artificial vulnerabilities, we analyzed the proportion of vulnerabilities patched in both categories. \textbf{Figure~\ref{fig:plot_rq1}} presents a comparative visualization of patching rates for real and artificial vulnerabilities across different models.

\textbf{Table~\ref{tab:patching_rates}} presents the number and percentage of real and artificial vulnerabilities patched by each model. The results show that while LLMs successfully patch a subset of real vulnerabilities, their effectiveness in patching artificial vulnerabilities is consistently lower across all models, indicating a challenge in addressing these cases. Moreover, we found that, while some models perform better than others, none maintains consistent performance across both categories. All models show a noticeable drop in effectiveness on the artificial vulnerabilities. The gap between their effectiveness in patching real and artificial vulnerabilities suggests that artificial vulnerabilities pose a greater challenge for these models.

\begin{quote}
\begin{adjustwidth}{-0.5cm}{-0.5cm}
\textbf{RQ2: How complementary and overlapping are LLMs in patching real vulnerabilities vs. artificial vulnerabilities?}
\end{adjustwidth}
\end{quote}
To investigate how LLMs complement or overlap in their patching capabilities, we analyze the extent to which multiple models patch the same vulnerabilities vs. cases where a single model patches a vulnerability that no other model does.

\begin{figure*}[tp]
    \centering
    \includegraphics[width=\linewidth]{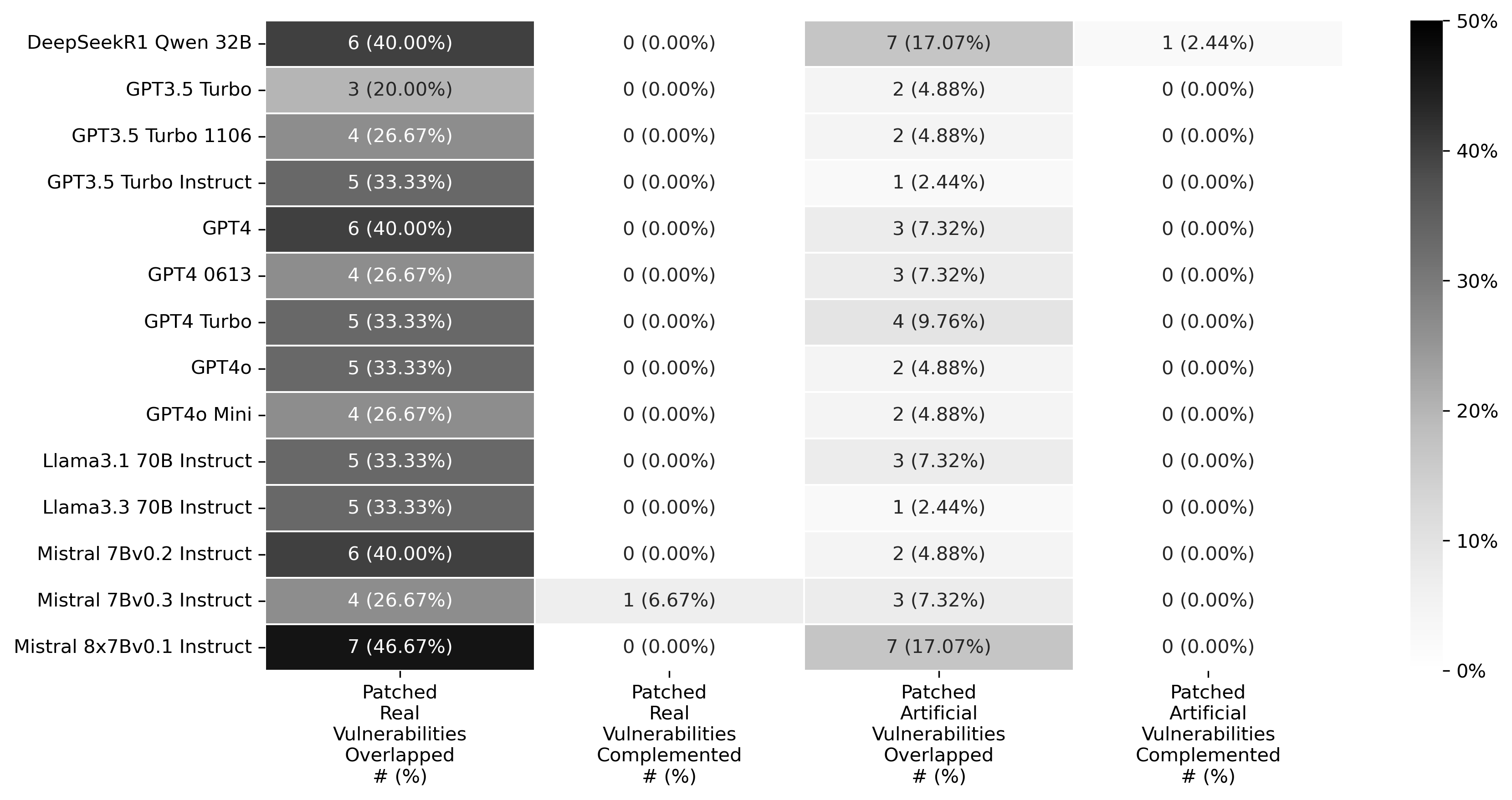}
    \vspace{-2em}
    \caption{Overlapping and complementarity of LLMs in patching real and artificial vulnerabilities}
    \vspace{-1em}
    \label{fig:plot_rq2}
\end{figure*}

\textbf{Figure \ref{fig:plot_rq2}} presents a greyscale heat-map in which each cell is annotated as X (Y\%), where X is the absolute number of vulnerabilities a given model patched in that category and Y is the corresponding share of that model’s total patches. The figure presents our drawn observations mentioned below:
\begin{enumerate}
    \item \textbf{Substantial overlap on real-world vulnerabilities.} All models patch at least three of the fifteen real CVEs in common with another model, and the most overlapping, i.e., \textit{Mistral 7B}, \textit{DeepSeekR1 Qwen 32B}, and \textit{GPT4}, overlap on 6-7 real CVEs (40-46.7\% of their respective real-vulnerability fixes). Hence, when a real vulnerability is within reach of current LLMs, several models generally converge on the same repair.
    \item \textbf{Marked reduction in overlap on artificial variants.} For the 41 artificial vulnerabilities, the overlap count per model falls to the 1–4 range for most LLMs (2.4\%–9.8\%), with only \textit{DeepSeekR1 Qwen 32B}, and \textit{Mistral 8x7Bv0.1} reaching seven overlaps each (17\%). Agreement among models is therefore far lower on artificial vulnerabilities, confirming the difficulty already highlighted in RQ1.
    \item \textbf{Scarcity of truly unique (complementary) patches.} Complementary successes, i.e., cases where a vulnerability is fixed by one model and by no other, are almost absent. Across all 56 vulnerabilities and 14 models, only two such instances appear: a single real vulnerability uniquely patched by Mistral 7Bv0.3 Instruct and one artificial vulnerability uniquely patched by DeepSeekR1 Qwen 32B.
    \item \textbf{Implications for ensemble strategies.} The combination of high overlap on real CVEs and negligible complementarity indicates limited marginal benefit from ensembling the considered LLMs in our study. The results indicate that most gains would come from whichever single model already attains the highest individual patching performance, rather than from additive coverage.
\end{enumerate}

Collectively, these findings indicate that contemporary LLMs tend to \textbf{pile up} on the same and relatively tractable known vulnerabilities, and mostly \textbf{fail} to patch the artificial ones, with a seldom exception of unique patches that other LLMs miss.

\section{Threats to Validity}
\textbf{Construct Validity.} The threat to construct validity lies in the measurement of patch effectiveness. We used \textit{PoV} test cases provided by Vul4J\cite{BuiSF22}, which may not comprehensively capture all security implications of the vulnerabilities. As a result, passing PoV tests does not necessarily imply complete vulnerability mitigation. Consequently, a patch that passes the existing PoV tests may still leave latent attack vectors unexercised, resulting in false negatives. Notably, \textit{Wang et al.}~\cite{wang2025vulnrepairevalexploitbasedevaluationframework} report that even the best LLM addressed \textit{only 5 out of 23 real CVEs} (approximately 21.7\%) under strict exploit-based validation. This suggests our results are consistent with low LLM patching rates found in other settings.

Another threat lies in the absence of a direct comparison with traditional or domain-specific patching solutions. We did not integrate baseline tools such as classical APR frameworks or static-analysis-driven fixers when evaluating the LLMs. This was intentional to maintain our study's focus on how different LLM-based approaches compare to one another, leaving cross-paradigm comparisons for future work.

\textbf{Internal Validity.} The threat to internal validity involves potential variations in LLM-responses due to their inherent stochasticity. To mitigate this, we used a consistent and standardized prompt across all experiments and evaluated each LLM independently. Running multiple generation trials per vulnerability could further mitigate randomness, but our single‑prompt, single‑run setup reflects a realistic one‑shot usage scenario.

Another threat lies with our limited prompting strategy. Since our evaluation employs a standardized single prompt, this may not fully leverage the LLMs’ potential. More sophisticated prompting strategies could improve patch-generation success and overall model performance. We chose not to explore those variations here in order to keep the evaluation consistent across models, leaving a study on advanced prompting techniques as future work.

\textbf{External Validity.} The threat to external validity involves the selection and representativeness of vulnerabilities included in our study. Although we considered a diverse set of CVEs and artificial vulnerabilities, our results might not generalize to other types of vulnerabilities or different scenarios.

\section{Conclusion and Future Work}
In this paper, we presented an empirical evaluation of Large Language Models (LLMs) in automated vulnerability patching. Our analysis, covering real and artificial vulnerabilities, highlights key insights into the capabilities and limitations of these models. Our results indicate that while LLMs demonstrate considerable effectiveness in patching real vulnerabilities, their performance significantly drops when addressing artificial vulnerabilities. This difference underscores a critical gap in their current ability to generalize patching strategies beyond previously encountered vulnerability patterns.

Additionally, our findings reveal that patching overlap among LLMs is notably higher for real vulnerabilities compared to artificial ones. This suggests that artificial vulnerabilities, despite exhibiting the same failure patterns, pose unique challenges that result in diverse patching outcomes across different LLMs.

In the future, we envision several avenues to extend this research. First, we plan to explore multi-model ensembling for patch generation, e.g., combining the suggestions of multiple LLMs or multiple independent executions to increase patch correctness. An ensemble could vote on or cross-verify patches, potentially overcoming individual model limitations, e.g., by voting or consensus among multiple LLM outputs. Secondly, we aim to investigate prompt tuning for this task. By tuning the prompts, we may significantly improve the models’ understanding of security fixes. Such tailored prompting could reduce errors that occur in single-prompt mode. Thirdly, we would like to expand the dataset to include a larger and more diverse set of vulnerabilities, covering additional vulnerability categories and languages (beyond those in the current \textit{Vul4J} subset). This will require reliable Proof-of-Vulnerability (PoV) tests for every new case, which is a labor-intensive manual process involving a thorough analysis of the vulnerable program, its human patch, and crafting an exploit that fails only when the fix is correct. Creating these artifacts is beyond our current scope, yet it remains an important next step for a broader generalization. Finally, exploring a combination of approaches like integrating static analysis with LLM suggestions or an interactive human-in-the-loop patch validation is an exciting direction to increase the practicality of LLM-based vulnerability patching.


\bibliographystyle{splncs04}
\bibliography{references}

\end{document}